\documentstyle[aps,epsf]{revtex}  

\begin{document}        

\baselineskip 14pt
\title{Triggering BTeV}
\author{Erik Gottschalk}
\address{Fermi National Accelerator Laboratory}
%
\maketitle              

\begin{abstract}        

BTeV is a collider experiment at Fermilab designed for precision studies of CP
violation and mixing. Unlike most collider experiments, the BTeV detector has a
forward geometry that is optimized for the measurement of $B$ and charm decays
in a high-rate environment. While the rate of $B$ production gives BTeV an
advantage of almost four orders of magnitude over $e^{+}e^{-}$ $B$ factories,
the BTeV Level 1 trigger must be able to accept data at a rate of 100 Gigabytes
per second, reconstruct tracks and vertices, trigger on $B$ events with high
efficiency, and reject minimum bias events by a factor of 100:1. An overview of
the Level 1 trigger will be presented.

\
\end{abstract}   	

\section{Introduction}               

The BTeV experiment at Fermilab is expected to begin running in the new
Tevatron C0 interaction region by the year 2005. The physics goals include
studies of CP violation and mixing, rare decays, and high sensitivity searches
for decays forbidden within the Standard Model. The primary focus of BTeV is on
precision studies of CP violation and mixing in $B$ decays.

BTeV benefits from the new Fermilab Main Injector, which was built to achieve
higher luminosity in the Tevatron. With a luminosity of $2\times 10^{32}
cm^{-2} s^{-1}$ the Tevatron can produce $4\times 10^{11}$ $B$ hadrons in
$10^{7}$ seconds of running. This rate of $B$ production is almost four orders
of magnitude larger than the $B$ production rate anticipated
for  $e^{+}e^{-}$ colliders operating at the $\Upsilon (4S)$ resonance.

In this paper I describe the Level 1 trigger for BTeV. I begin with an overview
of the BTeV detector and the operating environment in the C0 interaction
region. I describe the baseline design of the silicon pixel vertex detector,
which provides the data for the Level 1 trigger. The Level 1 trigger performs
track and vertex reconstruction to select events with detached vertices. The
goal is to trigger on $B$ decays with high efficiency, while rejecting minimum
bias (light quark) events.

Details of the baseline design of the Level 1 trigger have been published
\cite{walter}. Here I provide an overview of the
Level~1 trigger, report on
results from trigger simulations, and present some new ideas for track and
vertex reconstruction algorithms that challenge our baseline design. Our design
for both the trigger and the vertex detector will undoubtedly evolve as we
refine our understanding of detector hardware and physics simulations.

\section{The BTeV Detector}

BTeV is
optimized for $B$ physics.
The detector is a two-arm forward-geometry spectrometer (see
FIG. 1) designed to run at a luminosity of $2\times 10^{32} cm^{-2} s^{-1}$ and
a production rate of $4\times 10^{11}$ $B$ hadrons per year. This rate of $B$
production is high. However, the background from light quark events is also
large, with only 1 in 1000 events expected to be a $B$ event. To select a broad
spectrum of $B$ events efficiently,
BTeV will reconstruct tracks and vertices with a Level 1 trigger that
receives data from a state-of-the-art vertex detector. The Level 1 trigger
selects $B$ decays by reconstructing primary-interaction vertices, and by
identifying tracks that are \emph{not} associated with a primary vertex. The
goal is to trigger on tracks that come from $B$ decays, which are found as
secondary vertices detached from a primary vertex.

The trigger and vertex detector operate in a high-rate hadron-collider
environment. Like all of the new collider experiments in the Tevatron, BTeV is
being designed to operate with a Tevatron bunch spacing of 132 ns. Unlike most
hadron collider experiments, BTeV operates in a region with high track density,
due to the forward geometry. BTeV also operates close to the
Tevatron beams (the innermost edge of the vertex detector is within 6 mm of the
beams), and design considerations for both the vertex detector and the Level 1
trigger involve studies with beam luminosities in excess of
$2\times 10^{32} cm^{-2} s^{-1}$. At $2\times 10^{32}$ the mean number of
interactions per beam crossing is expected to be 2. Although events with two or
more interaction vertices could pose a problem for a Level 1 trigger designed
to trigger on detached vertices, the BTeV trigger benefits from a long
interaction region with $\sigma_{z}\,\approx \, 30\, {\rm cm}$, and is designed
to be relatively insensitive to multiple interactions per beam crossing.
Simulations show that the Level 1 trigger selects less than
1$\%$ of minimum bias interactions, even up to a luminosity of $2.5\times
10^{32} cm^{-2} s^{-1}$.


\begin{figure}[ht]	
\centerline{\epsfxsize 7.0 truein \epsfbox{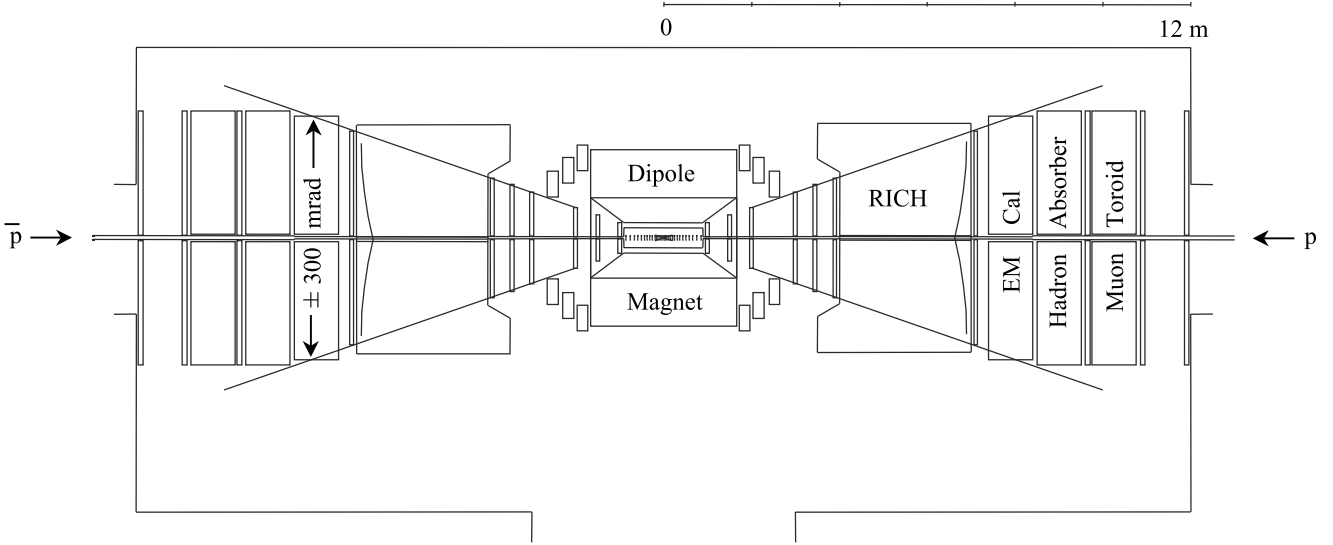}} 
\vskip +0.2 cm		
\caption[]{
\label{PlanFigure}
\small Plan view of the BTeV detector in the C0 interaction region.}
\end{figure}


The two-arm forward-geometry design of the BTeV spectrometer is accompanied by
significant design challenges. However, the forward geometry provides
considerable advantages for $B$ physics \cite{acceptance}, compared to a more
central detector. The main advantage is a larger Lorentz boost, which increases
the reconstruction efficiency for $B$ decays and improves the proper time
resolution. Another significant advantage is that a forward spectrometer can
have a much longer detector volume, and we exploit this aspect of the
geometry by including state-of-the-art particle identification for hadrons.
Furthermore, having a two-arm
configuration with a dipole magnet centered on the interaction region means
that BTeV has two spectrometers, thereby doubling the acceptance for $B$
physics compared to a single-arm spectrometer. As a double-arm spectrometer,
BTeV covers both the forward and the backward rapidity regions with a combined
coverage of $1.5<|y|<4.5$.

BTeV includes detectors (see FIG. 1) featured in many collider experiments,
with inner and outer tracking systems, E\&M calorimetry, and muon detection.
BTeV also includes a Ring Imaging Cherenkov (RICH) counter in each arm of the
spectrometer. The RICH detectors are ideal for charged-hadron
identification, so that different types of $B$ decays can be distinguished from
one another. The spectrometer has inner and outer tracking systems that
are highly segmented.
The inner tracking system, used for precision tracking and vertex
reconstruction, is the ``centerpiece'' of the BTeV spectrometer, and it consists
of planar pixel arrays located inside the Tevatron beam vacuum. FIG. 2 shows a
close-up view of 13 out of a total of 31 inner tracking stations. In our
baseline design the inner tracking system consists of silicon pixel detectors;
however, diamond pixel detectors are also being considered.

The inner tracker/vertex detector is a key component of the
BTeV spectrometer. There are 31 pixel tracking stations. In the central part of
the vertex detector the tracking stations are separated by 3.2 cm
(center-to-center); at the outer ends of the vertex detector the spacing for 8
of the 31 tracking stations is increased, and the total length of the detector
is 128 cm. Each station has three pixel planes that are arranged in views
\cite{views} with respect to the magnetic field. There are two bend views and
one non-bend view. Each pixel plane has over 500,000 pixels in a 10 x 10 cm
area, excluding the beam region. Each pixel provides an X and Y position
measurement, and a pulse-height measurement. The pixels, which are 50 x 400
$\mu$ in size, are arranged on sensor chips that are tiled to provide close to
100 $\%$ coverage over the active area of the vertex detector. FIG. 3 shows a
diagram of sensor chips for three quadrants of a pixel plane, and shows the 1.2
x 1.2 cm beam hole at the center of the vertex detector \cite{pixel}.
The beam hole is
larger during injection of the Tevatron beams, and is brought into the
configuration shown in FIG. 3 after the beams have stabilized.

\section{Level 1 Trigger}

The Level 1 trigger selects events by detecting the presence of $B$
decays. These decays are detected by first reconstructing primary interaction
vertices. Tracks from $B$ decays are then found with an impact
parameter cut that selects the tracks that miss the primary vertex by a
significant amount. To accomplish these tasks quickly, the Level~1 trigger
hardware receives data directly from the pixel detectors (at a rate of 100
Gb/sec). The trigger itself consists of three stages. The first stage is the
\emph{segment finder}, where hits from the three pixel planes per tracking
station are assembled into 3-dimensional space points, each with a
track-direction mini-vector. These mini-vectors are used for track
reconstruction in the second stage of the Level 1 trigger. In the third stage,
the reconstructed tracks are used to reconstruct primary interaction vertices,
calculate impact
parameters, and select tracks coming from secondary vertices.

The Level 1 trigger is heavily pipelined throughout the 3-stage reconstruction
process, and performs many operations in parallel to accommodate the data flow
from the pixel detectors. Data from an individual pixel consists of an X and Y
position measurement, a pulse height measurement, and a time stamp (in units of
132 ns) that is used by the trigger to assemble all of the data belonging to a
particular beam crossing. Data from adjacent pixels are combined into a pixel
\emph{hit} by a clustering algorithm that uses pulse height and position
measurements. Pixel hits are processed in parallel, for each pixel station, by
the \emph{segment finder}. Additional parallelism is obtained by subdividing
pixel hits into $\phi$ slices (see Section IV). The hits for each station are
combined into groups of three hits (one from each of three pixel planes), and
are used to calculate a 3-dimensional space point and a track-direction
mini-vector. FIG. 2 shows a schematic representation of the mini-vectors found
for a simulated $B^{o}\rightarrow \pi^{+}\pi^{-}$ event. The red line segments,
which appear to penetrate each tracking station, represent the mini-vectors
that are found by the first stage of the Level 1 trigger.


\begin{figure}[ht]	
\centerline{\epsfxsize 7.0 truein \epsfbox{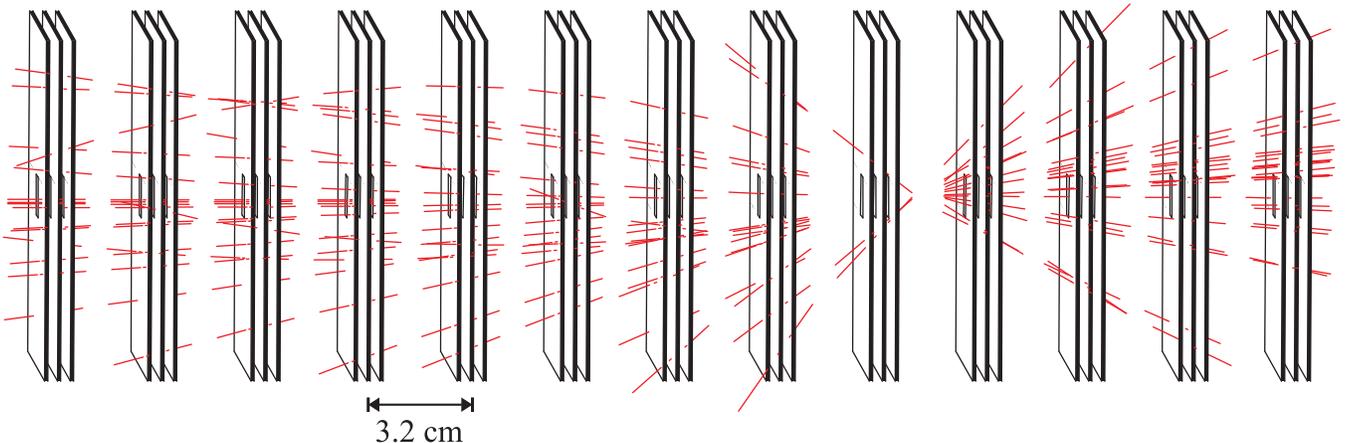}}   
\vskip +0.3 cm		
\caption[]{
\label{CloseFigure}
\small Close-up view of 13 out of a total of 31 pixel tracking stations,
which are located inside the Tevatron beam vacuum.}
\end{figure}


In the second stage of the trigger, the slope and position measurements for
each mini-vector in a pixel station are used to find matching mini-vectors in a
neighboring station. Mini-vectors with compatible measurements are combined to
form particle trajectories. In the third stage of the trigger, the curvature of
these reconstructed trajectories is used to eliminate low momentum tracks,
which tend to have large multiple Coulomb scattering errors. We also eliminate
tracks with large impact parameters (greater than 2 mm, for example). This
reduces the number of incorrect trigger decisions caused by tracks associated
with other interactions
that occurred during the same beam crossing. The decision to keep
or reject the data for a particular beam crossing is based on the number of
tracks that are found with a normalized impact parameter ($b/\sigma_{b}$)
greater than some value. For example, the requirement that there be at least 2
tracks with $b/\sigma_{b} > 3.5$ yields a trigger efficiency
of 40$\%$ for
$B^{o}\rightarrow \pi^{+}\pi^{-}$ events \cite{efficiency},
while providing a rejection factor of
$5\times 10^{-3}$ for minimum bias events. This result comes from a simulation
of the full pattern recognition with simulated pixel hits, and is presented in
greater detail elsewhere \cite{joel,sheldon}.


\begin{figure}[ht]	
\centerline{\epsfxsize 7.0 truein \epsfbox{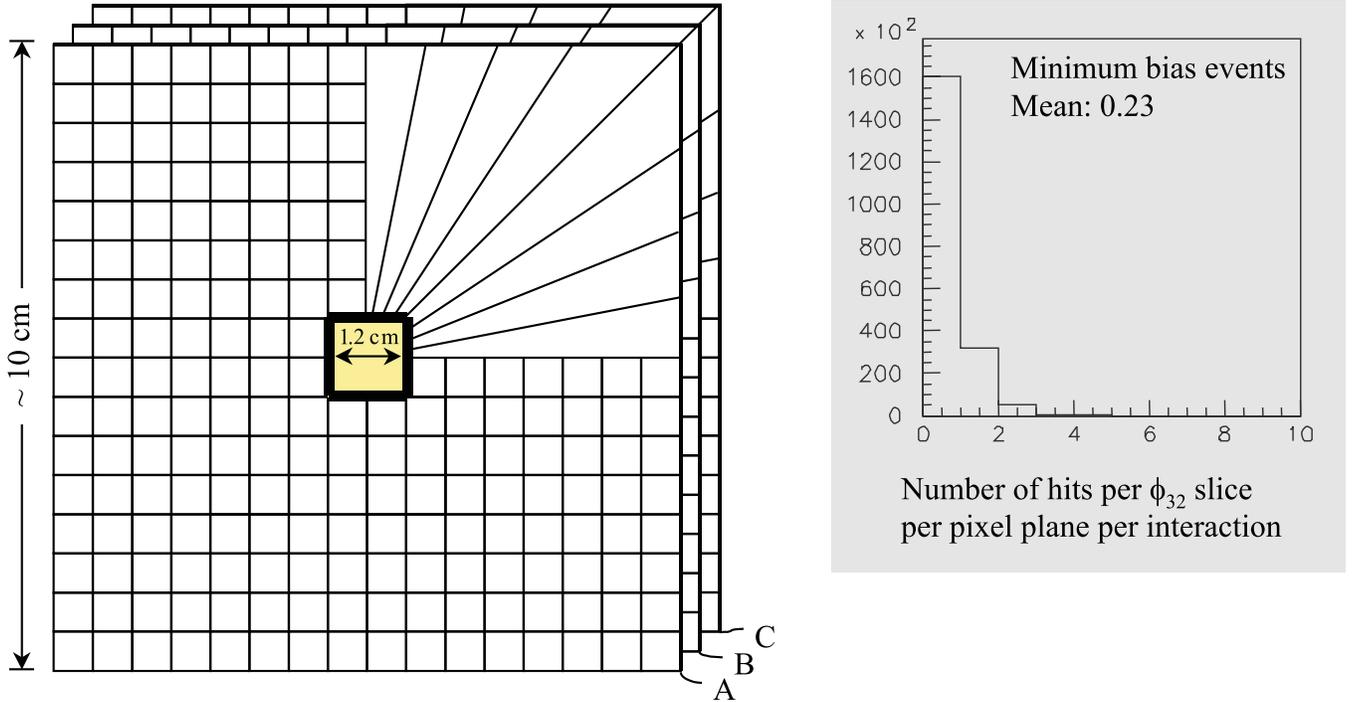}}   
\vskip +0.2 cm		
\caption[]{
\label{TrigFigure}
\small Trigger view of one quadrant (subdivided into eight $\phi$ slices)
in a pixel station with planes A, B, and C. The histogram on the right
shows the number
of hits for each plane per $\phi$ slice for minimum bias events.}

\end{figure}


\section{Trigger Hardware and Simulations}

With 100 gigabytes of pixel data coming into the Level 1 trigger every second,
a beam crossing every 132 ns, and an average of 2 interactions per crossing,
the time required for each pipeline step in the trigger is critical. Many of
the timing issues have been studied by performing a variety of trigger
simulations. These include Monte Carlo studies of the pattern recognition and
reconstruction algorithms, and simulations of the trigger hardware. The trigger
hardware and software algorithms will be extensively tested with a trigger
prototype that is being built with components specified in our baseline trigger
design. The design utilizes FPGAs (field programmable gate arrays) and two
types of DSPs (digital signal processors): a fixed-point DSP, and a
floating-point DSP. The fixed-point DSP is made by Texas Instruments and
belongs to the TMS320C6X family of DSPs. The floating-point DSP is an Analog
Devices ADSP-2106x SHARC. The layout of the logic for the FPGAs is mostly
complete. We are currently developing the software algorithms (including
optimizations of the code) for the DSPs. I should note that future
implementations of the Level 1 trigger may use different DSPs (as new DSPs with
better performance are introduced), and that our DSP algorithms will
undoubtedly be modified as we optimize the performance of the trigger.

Simulations of the Level 1 baseline design show that the first stage of the
trigger, the segment finder, requires the most computational power compared to
other parts of the trigger. A large number of operations are performed by the
segment finder, due to the number of combinations of pixel hits that must be
sampled to find the hits that define a mini-vector. To address this problem, we
introduce considerable \emph{parallelism} in the trigger architecture. This is
a key feature of the baseline trigger design. The parallelism begins with the
organization of the data read out from the pixel detectors. Pixel hits are
read out in parallel for each quadrant of a pixel plane. The hits from the three
planes that belong to a pixel station are brought together in a \emph{quadrant
processor}, which represents one of 124 identical circuit boards that make
up the first stage of the trigger. Additional parallelism is introduced by
subdividing the pixel hits in a quadrant into 8 $\phi$ slices, for a total of
32 $\phi$ slices per pixel station. The hits in each $\phi$ slice are
processed in parallel by one of the 992 TMS320C6X DSPs (a total of 992 DSPs for
31 pixel stations, each with 32 $\phi$-slice DSPs).
FIG. 3 shows the ``trigger view'' of one quadrant, and shows that
the subdivision of hits into $\phi$ slices reduces the mean number of hits
per pixel plane to a value of 0.23 hits for minimum bias events. This is a
small number of hits (on average), and with suitable buffering in the Level 1
trigger we can average over numerous beam crossings in our timing studies.

Even with the small number of hits per $\phi$ slice, the timing for the
segment finder is critical. Table I shows timing results for the segment finder
algorithm developed for the TMS320C6X DSP. In the table
we define four cases to present our results. For a given $\phi$ slice and
with three planes (A, B, and C) in a pixel station (as shown in FIG.~3), we
have a single hit in each plane for one track passing through a pixel station.
This is the (1,1,1) case in Table~I. For two tracks passing through the same
$\phi$ slice we have two hits per plane, or (2,2,2). With three and four
tracks in the same $\phi$ slice we get the (3,3,3) and (4,4,4) cases,
respectively. The timing results for the four cases are divided into
three categories that represent different levels of optimization for the
segment finder algorithm. The first category involves $C$ code that has
\emph{not} been optimized (the code is compiled without any optimization), and
we use this to establish a reference measurement for subsequent measurements
involving different levels of optimization. For the second category we allow
the $C$ compiler for the TMS320C6X DSP to optimize the algorithm, and for the
third case the programming is done directly in assembly language. Although
programming in assembly language is more cumbersome, we can optimize the code
so that all hardware components in the DSP are performing useful operations
most of the time. Our timing results in Table I show that the $C$ compiler for
the TMS320C6X DSP is not very effective at optimizing the segment finder
algorithm; the compiler achieves only a minor performance gain for the (1,1,1)
case, and at best a factor of two improvement for cases with more hits in a
$\phi$ slice. Table I also shows that we are able to achieve an order
of magnitude improvement \cite{notquite} in the timing by optimizing the
algorithm in assembly language.

To obtain an estimate of the average time required to find mini-vectors in
minimum bias events, we compute a weighted average by using the results in
Table I weighted by the distribution of hits in FIG. 3. We use a value of zero
for the (0,0,0) case, since the DSP does not perform any operations
when there are no hits in a $\phi$ slice. The average time for each level
of optimization is shown in the last row of Table I. Although these results are
somewhat oversimplified, they do indicate that an assembly language
implementation of the segment finder algorithm is necessary 
so that the first stage of the trigger stays within a
budget of 132 ns per beam crossing. The average time of 37 ns for the
segment finder is encouraging, since it suggests that we are close to having a
feasible Level 1 trigger design. We continue to work on more refined
simulations of the trigger, and on developing the trigger prototype
for detailed studies of the trigger hardware.

 \begin{table}
 \caption{Timing results in nanoseconds for the segment finder algorithm
 running on the TMS320C6X DSP.}
 \begin{tabular}{cccc} 
 Number of hits  &Time for  &Time for compiler  &Time for assember\\
 per $\phi$ slice in  &non-optimized  &optimized  &optimized\\
 planes (A,B,C)  &segment finder(ns)  &segment finder(ns)  &segment finder(ns)\\
 \tableline 
 (1,1,1)  &1915  &1630  &135\\
 (2,2,2)  &5705  &3670  &315\\
 (3,3,3)  &14015  &7560  &1340 \cite{notquite}\\
 (4,4,4)  &28915  &13840  &2885 \cite{notquite}\\
 \tableline 
 weighted average  &530  &395  &37\\
 \end{tabular}
 \end{table}

\pagebreak

\section{Alternatives to the Baseline Design}

Although most of our trigger-design efforts are devoted to the development of
the baseline design of the Level 1 trigger, we are also exploring alternative
designs.
These alternatives invariably involve design changes in both the
Level
1 trigger and the vertex detector, since the two systems are interdependent.
For example, we are investigating a design that entails two pixel planes per
tracking station instead of the three-plane tracking stations in our baseline
design. The two-plane design for the vertex detector has less material, which
reduces multiple Coulomb scattering errors. Other advantages include a
reduction in the heat load for the pixel cooling system, and perhaps a
reduction in cost. These improvements in the vertex detector require
significant changes in the Level 1 trigger. The segment finder in the baseline
design must be replaced by a different algorithm to initiate the
track-reconstruction phase of the trigger. We are working on algorithms that
find track segments spanning three tracking \emph{stations}, compared to the
three-plane mini-vectors in the baseline design. In one approach, we are
investigating a massively parallel system based on an FPGA design that can
handle the large number of pixel-hit combinations that must be sampled
to identify three-station
track segments. In a second approach, we reduce the number of pixel-hit
combinations
that must be sampled by finding the first three
pixel hits for each track as it passes from the beam region into the vertex
detector.
In this approach each track is found once, and is then projected to neighboring
stations to extend the track and improve the momentum measurement for the track.
Both of the alternative trigger designs require additional work before they can
be considered as viable alternatives to our baseline trigger design.

\section{BTeV Status}

BTeV is an approved R\&D project. The goal of this
project is to conduct all detector R\&D, and to design the BTeV experiment.
Although the forward-geometry of the BTeV detector offers numerous design
challenges,
the benefits of a second-generation experiment dedicated to 
\emph{precision} studies of $B$ physics are
substantial. A technical design report will be submitted in 14 months.


\begin{references}
\bibitem{views} We define a ``view'' for a pixel detector as being analogous to
 the orientation of a silicon strip detector. Since the pixels are rectangular
 in shape, the long dimension of a pixel is analogous to the long dimension of
 a silicon strip.
\bibitem{efficiency} The trigger efficiency is calculated after requiring that
 both pions are in the spectrometer and accepted for further analysis.
\bibitem{notquite} The assembly language implementation for the (3,3,3)
 and (4,4,4) cases has not been fully optimized, so an additional improvement
 in speed is achievable for these cases.
\end{references}
\end{document}